**Topological Metamaterial for Magnetic Resonance Imaging**

Siyong Zheng[1,&], Maopeng Wu[2,&], Zhonghai Chi[1], Xinxin Li[2], Mingze Weng[2], Fubei Liu[1], Yingyi Qi[3], Yi Yi[4], Yakui Wang[4], Jie Gao[4], Guoxiang Zhan[3], Zewen Chen[5], Shuojun Ling[2], Yucheng Wei[1], Zhuozhao Zheng[4], Qian Zhao[2*], Ji Zhou[1*]

[1]State Key Laboratory of New Ceramics and Fine Processing, School of Materials Science and Engineering, Tsinghua University, Beijing, China
[2]State Key Laboratory of Tribology in Advanced Equipment, Department of Mechanical Engineering, Tsinghua University, Beijing, China
[3]Tsinghua Shenzhen International Graduate School, Tsinghua University, Shenzhen, China
[4]Department of Radiology Beijing, Tsinghua Changgung Hospital, School of Medicine, Tsinghua University, Beijing, China
[5]Weixian College, Tsinghua University, Beijing, China

[&]These authors contributed equally
*Corresponding authors, zhouji@tsinghua.edu.cn; zhaoqian@tsinghua.edu.cn

**Abstract** [Magnetic Resonance Imaging (MRI) is crucial in global healthcare, but the traditional receive coils, as a core component of MRI, SNR enhancement is limited due to the optimization of channel number and magnetic field strength faces high cost and complexity challenges. Here, we demonstrate the use of a topological material to enhance MRI signal reception. Designed with a stack of weak couplings, this material forms quasi-two-dimensional dual topological boundary states. High properties are achieved through low-loss signal transmission via these topological states, as well as only enhanced local magnetic fields and increased number of channels. Initial tests demonstrate superior performance and accessibility compared to commercial coils, suggesting significant potential. This concept introduces a transformative paradigm for all MRI coil designs.]

**Main**

Magnetic Resonance Imaging (MRI) is one of the most essential pillars in global healthcare[1-4]. Pursuing higher resolution and image quality is crucial for early disease diagnosis and revealing biological processes in vivo[5-7]. In the phenomenon of nuclear magnetic resonance, hydrogen nuclei in the body respond to the external radiofrequency (RF) excitation field ($B_1^+$), causing relaxation and emitting RF signal fields ($B_1^-$). These signals contain critical information about anatomical structures, tissue composition, and pathological changes[7].

MRI uses coils to receive RF signals ($B_1^-$) from the region of interest (ROI). Array coils are the most typical representatives in MRI, dominating clinical practice due to their tight coupling with the patient and local resonance enhancement of the $B_1^-$ field [8]. Array coils also support parallel imaging techniques, which accelerate k-space sampling by utilizing the spatial sensitivity of multiple coils, thereby significantly reducing scan time and improving imaging efficiency. However, array coils are expensive and lack cross-platform flexibility. Over the past 30 years, researchers have focused on increasing the number of channels to enhance the image signal-to-noise ratio (SNR) and improve parallel imaging capabilities[9,10], but this approach now faces severe decoupling issues [11-15]. To solve the coupling problems among numerous coils, an increasing number of non-magnetic electronic components, cable traps, baluns, adapters, and interfaces are added to array coils to meet the demands of extremely complex decoupling circuits. As a result, the performance of array coils has been in a dilemma, their cost-effectiveness has declined, and they have become increasingly bulky. Additionally, these coils often require dedicated interfaces and complex electronic decoupling networks, leading to deep integration with specific MRI manufacturers' systems. This design not only limits their interoperability across different systems but also increases the complexity of system integration, operation, and maintenance.

Over the past 20 years, metamaterials have been considered a potential alternative to array coils due to their ability to control electromagnetic wave propagation and field distribution, as well as their wireless and lightweight properties [16-29]. Initially, metamaterials in MRI used "Swiss roll" arrays to guide magnetic flux and transmit signals from areas like the thumb to distant coils. However, this design only enhances electromagnetic waves along the magnetic field axis, requiring the axis to be perpendicular to B0 and a height of up to 20 cm, which poses challenges for clinical use.

With deeper research, recent MRI metamaterials have enhanced $B_1$ field strength through tight coupling with the patient and local resonance, improving sensitivity [21,27,30-33]. This design operates similarly to traditional array coils, but the localized $B_1^-$ strength in metamaterials is nearing its limit, leaving limited potential for further enhancement, causing their performance to saturate.

As a result, research has increasingly focused on expanding the application scenarios and functionalities of MRI metamaterials, especially to improve patient comfort [28,29,34-36]. A notable advancement is the integration of metamaterials with local receive coils to enable parallel imaging [26].

Notwithstanding this apparent contradiction of MRI matematerials, metamaterials may continue to demonstrate their potential for enhancing both signal transmission fidelity and reception sensitivity in MRI systems [24]. Research on topological edge states offers new possibilities for overcoming the limitations of MRI metamaterials by enhancing the transmission and reception efficiency of MR scanners [24,37]. The SSH topological chain, constructed from coupled LC loops, can effectively localize the $B_1$ field. Its odd-numbered structure produces semi-infinite edge states, demonstrating unique advantages in thymus imaging[38-40]. Therefore, further enhancing the transmission and reception efficiency of topological edge states, improving image SNR, reducing their size and weight, and advancing their clinical versatility and accessibility holds significant value and presents considerable challenges.

Here, we propose an innovative design for receive coils and materials that combines long-distance signal transmission with localized $B_1$ field capabilities. This new material, based on the concept of topological insulators, integrates topological boundary states and chiral symmetry (Supplementary S7). We call this material Topological Magnetic Resonance Metamaterial (TMRM). TMRM demonstrates superior topological signal transmission and phase transition capabilities, acting as an efficient "movable bridge" that connects magnetic resonance signals to MR receive coils.

In its $B_1^-$ state, TMRM is designed to be interface-free, passive, and requires no special protocols, making it easy to integrate with existing MRI equipment. In clinical MR scanners, the built-in spine (SP) and body birdcage coils (BC) are commonly used but often suffer from weak SNR, with the SP also exhibiting poor field uniformity. By seamlessly integrating with these coils, TMRM directly enhances signal reception and improves image SNR. It naturally ensures a uniform $B_1$ field distribution, which results in clearer and more reliable images without the need for the complex post-processing algorithms required by conventional array coils.

When exposed to RF pulses, TMRM transitions into the $B_1^+$ state and undergoes a topological phase change into a trivial state—analogous to a movable bridge opening to allow large ships to pass. At this stage, TMRM does not amplify the $B_1^+$ field, thereby avoiding interference with image acquisition and ensuring patient safety.

Our initial demonstration on human wrist joint imaging, conducted on 10 healthy volunteers with a 1.5 T MR scanner, confirmed these advantages. TMRM significantly outperformed a standard four-channel flexible coil and delivered results comparable to a specialized 12-channel wrist coil across multiple MRI sequences. Notably, these improvements were achieved without any modifications to the MRI system or additional technician training. The enhanced SNR and intrinsic image uniformity provided by TMRM translate directly into improved diagnostic clarity and accuracy.

**Design Concepts and Magnetic Resonance Imaging Effects**

The design concept of TMRM is illustrated in Fig. 1a. TMRM utilizes topological boundary states to effectively guide the signal. Unlike intensely enhancing $B_1^-$ locally in region of interest (ROI), we use the weak topological insulator (WTI) theory to construct pseudo-two-dimensional dual topological boundary states (DTBS) [41,42]. Due to the unique characteristics of TMRM, the $B_1^-$ field is localized not only within the ROI but also in the outer region near the MR receive coil. This innovative field distribution enables TMRM to both sensitively receive signals from the ROI and effectively couple with the MR receive coil. Consequently, the induced current in MR receive coil is significantly strengthened. Furthermore, due to the topological energy transmission properties, TMRM exhibits minimal resistance loss, ensuring that the presence of the material does not increase signal attenuation as the signals propagate to the outer boundary states (Supplementary Fig. 10b). In contrast, traditional metamaterial coils provide only localized field enhancement, causing the $B_1^-$ field to weaken as it approaches the MR receive coil. As a result, there is weaker coupling between conventional metamaterial coils and the MR receive coil. metamaterial coils and the MR receive coil, limiting signal reception and current induction efficiency.

In this study, we evaluate the impact of Topological Magnetic Resonance Metamaterial (TMRM) on human wrist joint imaging in healthy volunteers. Ten volunteers were imaged in a prone position (Supplementary Fig. 6a). TMRM is designed to work in synergy with the built-in spine coil (SP) of the MRI system, thereby enhancing clinical applicability, accessibility, and flexibility. Although the spine coil is primarily intended for spinal imaging, it is typically unsuitable for joint imaging due to its uneven magnetic field distribution and limited long-range imaging capability. When used in conjunction with TMRM, these limitations are overcome as TMRM improves image signal-to-noise ratio (SNR) and uniformity while seamlessly integrating with standard MRI setups. This combined approach not only optimizes imaging quality but also broadens clinical applicability.

Using a 1.5 T MR scanner with standard clinical scanning parameters (Supplementary Table2), we compared human wrist joint imaging results acquired with TMRM, a standard commercial four-channel flexible coil (4-ch FLC), and a specialized 12-channel wrist coil (12-ch WR) across three commonly employed MRI sequences (Figure 1b): $T_1$-weighted spin-echo ($T_1$ SE), $T_2$-weighted fast spin-echo ($T_2$ FSE), and $T_2$-weighted multi-echo ($T_2$ GETI). The SNR is calculated solely from images (Supplementary S2). TMRM adapts seamlessly to multi-sequence scanning, delivering clear, high-contrast images suitable for clinical diagnosis. Under conventional scanning conditions, TMRM outperformed the 4-ch FLC and achieved performance comparable to the 12-ch WR (Fig 1c-e). Specifically, while the bone marrow SNR of TMRM is slightly lower than that of the 12-ch WR, the muscle and cartilage SNR are comparable, and overall image quality scores for TMRM were similar to those of the 12-ch WR and superior to the 4-ch FLC (Fig 1f).

In water phantom imaging (Supplementary Table3), when TMRM was used with the SP, the image SNR was comparable to that of the 12-ch WR (Fig. 1g). Conversely, when TMRM was combined with the BC, it outperformed the 4-ch FLC but remained inferior to the 12-ch WR, reflecting the higher sensitivity of the SP compared to the BC. Moreover, TMRM demonstrated superior image uniformity (Fig. 1h, Supplementary Fig. 25). Overall, the enhanced SNR achieved with TMRM resulted in clearer and more detailed images, thereby improving the accuracy of wrist condition diagnosis and assessment.

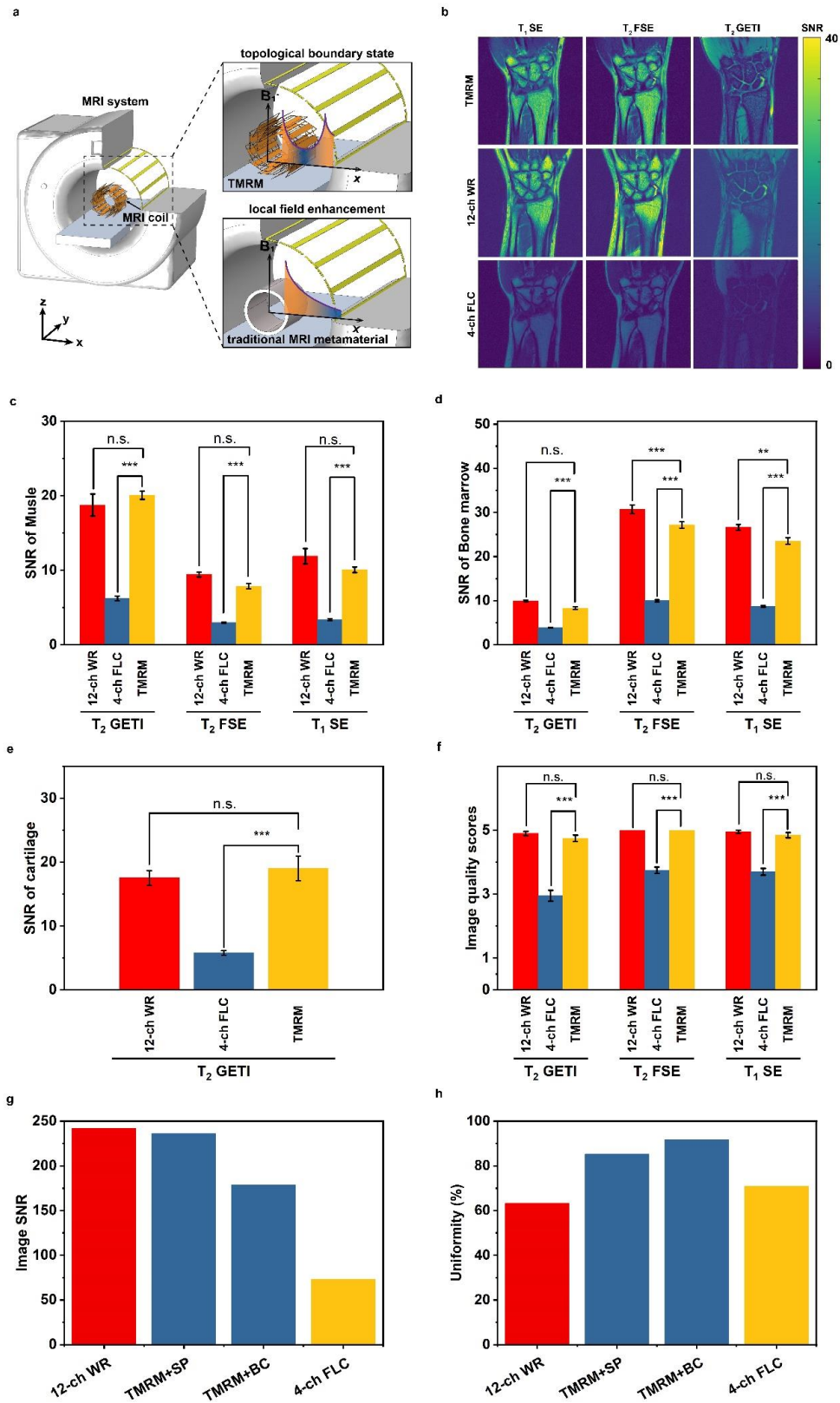

**Fig. 1. Concept and effect diagram. (a)** Schematic illustration of the principle behind using Topological Magnetic Resonance Metamaterial (TMRM) to improve MRI. **(b)** Image SNR maps of the human wrist joint under three sequences. **(c-e)** Image SNR analysis of wrist joint MRI scans of volunteers (n = 10). **(f)** Image quality scores of wrist joint MRI scans of volunteers (n = 10). **(g)** Image SNR analysis of the water phantom among TMRM, 12-ch WR, and 4-ch FLC. **(h)** Uniformity analysis of the water phantom among TMRM, 12-ch WR, and 4-ch FLC. All error bars represent standard error of mean (SEM); *$p < 0.05$, **$p < 0.01$, and ***$p < 0.001$. TMRM is the DTBS configuration.

**The Design and Simulation**

In this study, we aim to elucidate the process of achieving pseudo-two-dimensional dual topological boundary states (DTBS) in metamaterials and initiating topological phase transitions. While previous investigations into topological circuits, photonic crystals, and metamaterials have adeptly mimicked a variety of topological states observed in condensed matter models, their applicability to MRI remains limited[37,43-46]. The limitation stems from the reliance on voltage to emulate electronic energy levels in these studies, which is instrumental in manipulating near-electric field distributions. However, MRI technology prioritizes the enhancement of magnetic field strength and the minimization of electric field intensity. Moreover, fabricating topological materials that exhibit two-dimensional topological boundary states often entails considerable complexity and necessitates the use of magnetic components, which poses safety concerns in MRI environments. To simplify our demonstration and address these challenges, we have chosen to focus on the foundational Su-Schrieffer-Heeger (SSH) model (Fig. 2a)[47,48].

The design of TMRM is built upon the topological circuit. The topological circuit, which uses current to simulate the SSH model, is presented in Fig. 2b. By alternating the arrangement of impedances of two different impedances, we simulate the two types of electronic transition strengths and chiral symmetry in the SSH model. In reference to the language used in the tight-binding approximation model, we can write the impedance matrix of the circuit as:

$$\hat{Z} = j\omega L_v \sum_{m=1}^{N} (|m, Z_B\rangle\langle m, Z_A| + h.c.) + j\omega L_w \sum_{m=1}^{N-1} (|m+1, Z_A\rangle\langle m, Z_B| + h.c.). \tag{1}$$

Here, $Z_A$ and $Z_B$ represent the impedances of the lattice loops. In the matrix, the off–diagonal elements $j\omega L_v$ and $j\omega L_w$ represent the hopping strengths between lattice A and lattice B, proportional to the inductance values, while the diagonal elements $Z_A$ and $Z_B$ represent the on-site energies of quasi-particles on tracks A and B. Thus, this design mathematically corresponds to the SSH model.

To further substantiate the parallel between the described circuit and the SSH model, we employed numerical analysis on the impedance matrix. This approach allows us to illustrate the alterations in the circuit's band structure in response to variations in the inductance coupling strength, where modes with zero eigenvalues lie in the energy gap (Supplementary Fig. 8a). Using SPICE simulation software to examine the current distribution in the topological circuit loops showed that the current intensity was concentrated at both the initial and terminal loops (Supplementary Fig. 8b). This observation aligns with the expected distribution of the wave function for topological boundary states within the SSH model. Significantly, in this topological circuit, the loop impedance within the mesh emulates the energy levels typically associated with electrons. Thus, by modulating the parameters of the lumped elements, it becomes possible to engineer topological boundary states at varying frequencies. For instance, achieving a frequency of 63.8 MHz, which is congruent with the operational frequency of a 1.5 T MR scanner, demonstrates the practical applicability of our approach.

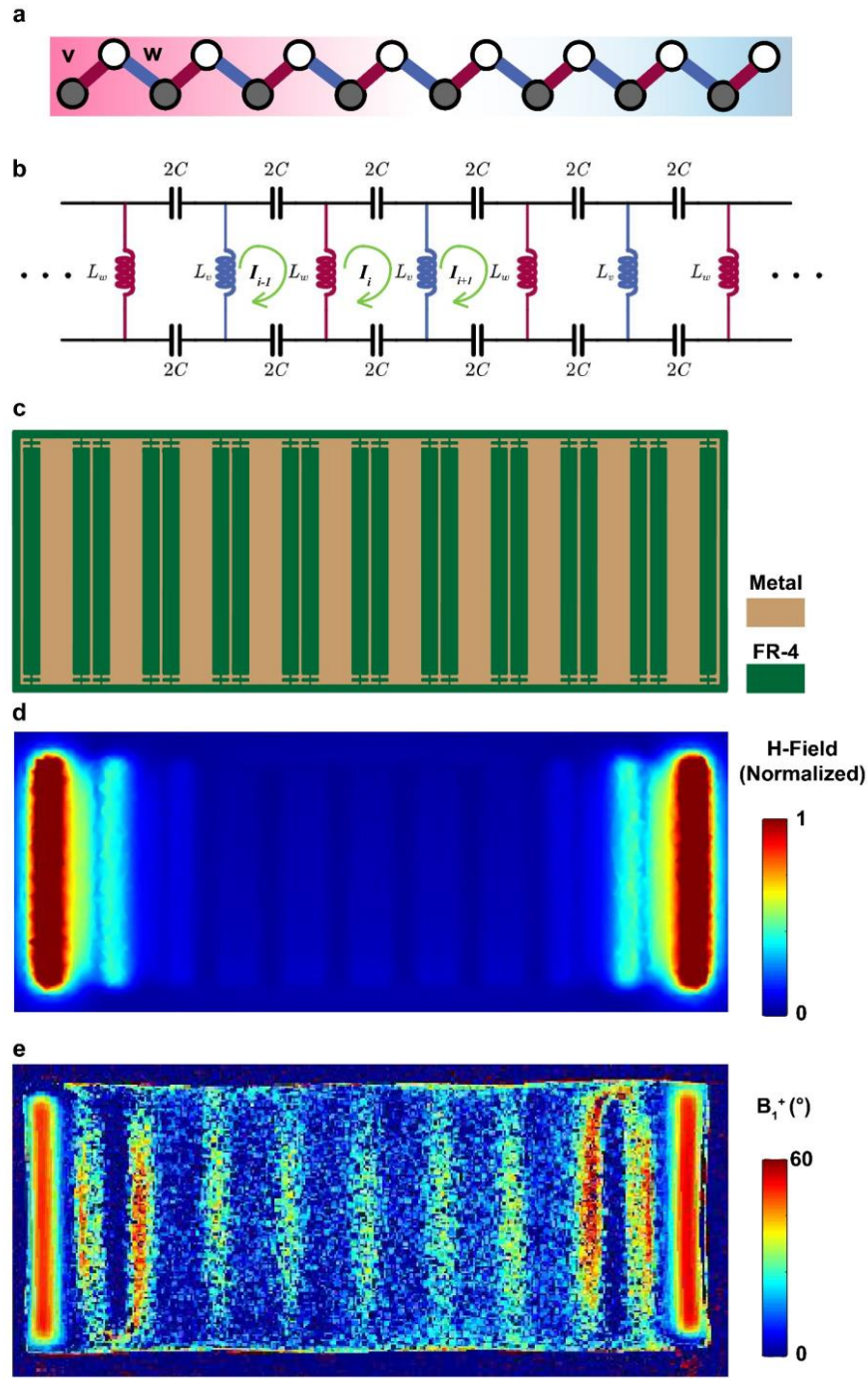

**Fig. 2. Design and Simulation of TMRM Sheets. (a)** Schematic illustration of SSH model. **(b)** A topological circuit. **(c)** The structure of TMRM unit. **(d)** A simulation diagram of the one-dimensional magnetic field distribution of the TMRM sheet. **(e)** $B_1^+$ maps in water phantom.

To determine if a topological circuit can influence the magnetic field distribution in mainstream MR scanners, the main challenge is transforming this abstract one-dimensional model into a tangible physical form. We used 82pF capacitors and rectangular metal strips (2mm and 20mm widths, serving as inductors) as physical counterparts for the alternating coupling in the SSH model. These elements were modeled in CST Microwave Studio (Fig. 2c) to analyze their impact on the near RF magnetic field. Results show that TMRM sheets can reproduce the SSH model's topological boundary states at the desired frequency (Fig. 2d). The TMRM unit was fabricated using standard commercial printed circuit board (PCB) techniques and consistent with simulation predictions at the target frequency (Fig. 2e). Using the standard Dual-Angle Method (Methods and Supplementary Table4) with a spoiled gradient echo sequence (SPGR), we tested the $B_1^+$ field in a water phantom and successfully observed that the TMRM sheets achieved one-dimensional topological boundary states.

## The Impact of Topological Boundary States

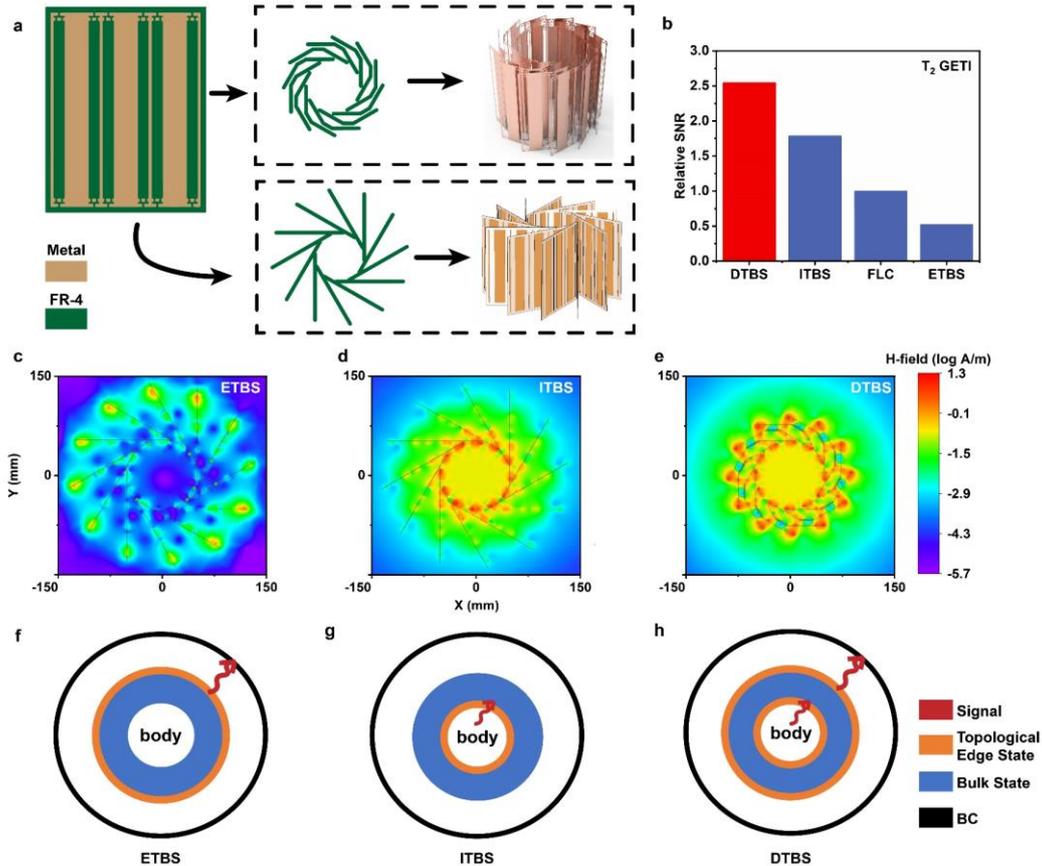

**Fig. 3. Design and Water Phantom Experiment of TMRM with Pseudo-two-dimensional Topological Boundary States. (a)** Schematic diagram of the stacking TMRM Sheets. **(b)** Bar chart of image SNR for water phantom under different topological boundary state conditions. **(c-e),** The impact of different topological boundary state distributions at 63.8MHz. **(f-h),** Diagram of the coupling relationship between the topological boundary states of TMRM, the magnetic resonance signals of human joints, and the BC of the MR scanner.

Accurate stacking of TMRM sheets is crucial for constructing quasi-two-dimensional dual topological boundary states. Inspired by the Weak Topological Insulator (WTI) theory, we use the electromagnetic coupling between the sheets to adjust the coupling between boundary states, demonstrating two stacking approaches (Fig. 3a).
One approach is to periodically and rotationally arrange the TMRM sheets along the tangential direction of a cylindrical wall. The uneven distance between the sheets, especially the extremely close proximity near the cylindrical wall, results in strong coupling. This design gives the TMRM boundary states non-degenerate energy levels, with the magnetic field localized only on one edge of the structure, forming quasi-two-dimensional semi-infinite topological boundary states. Thus, at the Larmor frequency, we can design materials with only external topological boundary states (ETBS) (Fig. 3c) or only internal topological boundary states (ITBS) (Fig. 3d). Another method involves adjusting the curvature angle of the TMRM sheets to regulate the coupling strength between the sheets. We meticulously designed the curved configuration of the TMRM sheets, ensuring their coupling is as weak, parallel, and consistent as possible compared to the unbent design. This achieves weakly coupled stacking of one-dimensional topological boundary states, forming quasi-two-dimensional dual topological boundary states (Fig. 3e). The magnetic field of the pseudo-two-dimensional dual topological boundary states is localized on both the inner and outer rings of the TMRM (Fig. 3e). The magnetic field localized on the inner ring uniformly covers a circular area with a diameter of 150 mm, while the magnetic field localized on the outer ring is positioned near the MR scanner's receive coil, the BC. The BC serves as both the MR scanner's RF transmit coil and can also be used as a receive coil. Typically, due to low SNR, the BC is not used as a receive coil, although it is very simple and convenient to use[49,50].

Notably, the proximity of the TMRM's outer ring magnetic field to the BC makes the TMRM extremely convenient to use, eliminating the need for additional interfaces or other receive coils that might come with the MR scanner (Fig. 3h). Using only the BC as a receive coil, the TMRM is expected to achieve high SNR imaging.

To compare the impact of single-sided topological boundary states versus dual topological boundary states on MRI imaging, we conducted MRI experiments on a water phantom using a commonly used 1.5 T MR scanner (Supplementary Table5). The TMRM was mounted on a fixed plastic support with a diameter of 150 mm, and a water phantom with a diameter of about 70 mm was placed at the center of the fixed support (Supplementary Fig. 6b). Different topological boundary states of the TMRM can affect the image SNR (Fig. 3b). Among them, the SNR of DTBS increased more than twofold compared to the FLC. ITBS exhibited a certain improvement in SNR. ETBS did not improve and even suppressed the image SNR. The performance of DTBS aligns with expectations, as it effectively guides the $B_1^-$ signals within the ROI and efficiently transmits the signals to the BC (Fig. 3h). Due to the topological boundary state characteristics of TMRM, there is minimal signal loss during transmission, resulting in a significant increase in SNR. Notably, the $B_1^-$ enhancement factor of ITBS within the ROI is identical to that of DTBS, and ITBS is very similar to volumetric metamaterials. However, compared to DTBS, the SNR enhancement is smaller (Fig. 3g) because the $B_1^-$ field of ITBS is weaker near the BC, leading to a reduced induced current in the BC. Although ETBS has a stronger coupling capability with the BC, it cannot effectively guide the magnetic field and couple the precession signal within the ROI. The signal transmitted by ETBS to the BC is nearly null, leading to the poorest SNR performance (Fig. 3f). Therefore, maintaining dual topological boundary states is crucial for MRI signal transmission and helps improve SNR.

## Adapting Sequences through Topological Phase Transition

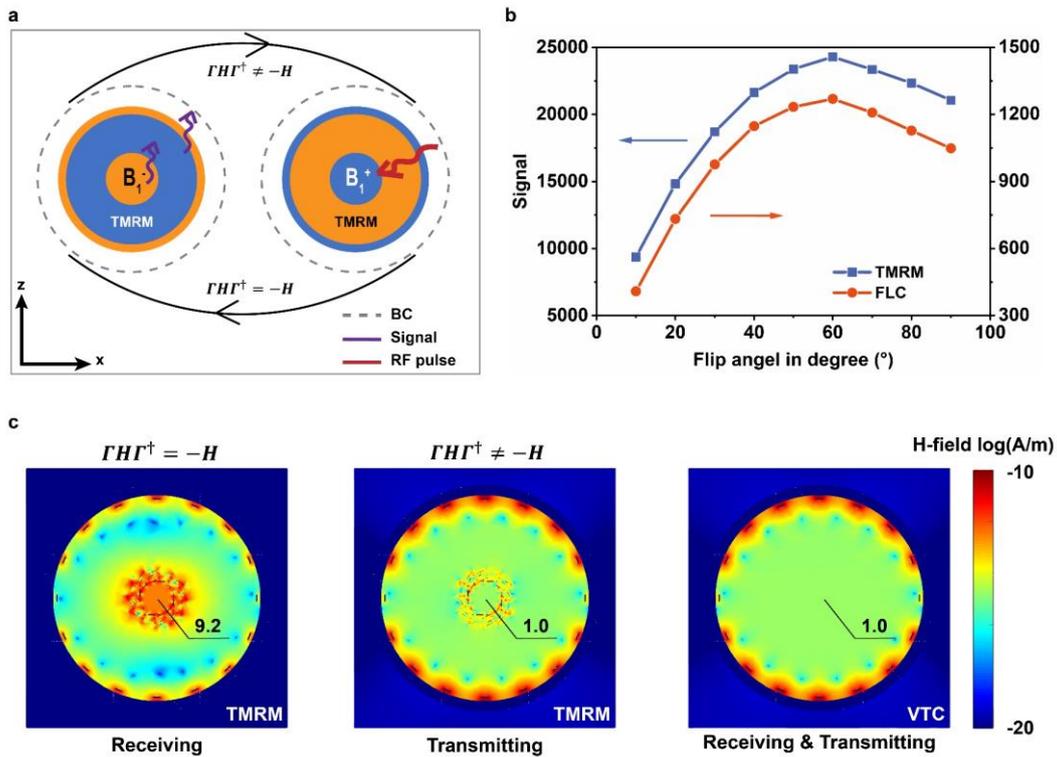

**Fig. 4. Effectiveness of Topological Phase Transition (a)** The topological phase transition process and chiral symmetry of TMRM. **(b)** Testing the relationship between signal and flip angle. **(c)** Magnetic field simulation illustrating the topological phase transition caused by maintaining and breaking chiral symmetry protection.

The adaptive capability of the TMRM based on topological phase transitions for different sequences is illustrated in Fig. 4. The TMRM, if without AC diodes, exhibits an identical magnetic field distribution under both $B_1^+$ and $B_1^-$ conditions, corresponding to the DTBS. To unpredictable deviation of the flip angle (FA) from the original settings in MRI sequences and a potential risk to patients [17-21], we exploit the nonlinear properties of AC diodes to introduce detuning in the TMRM edge circuits (Fig. 4a). During the transmission phase, high-power RF pulses cause the AC diodes to conduct (Supplementary S8), which activates the capacitors connected in parallel on the edge circuits and alters the circuit impedance, $Z_A$. Consequently, under $B_1^+$ excitation, this detuning disrupts TMRM's chiral symmetry, thereby triggering a topological phase transition. When in the radiofrequency transmission phase, the diodes conduct, and the impedance matrix of the topological circuit of the TMRM's unit can be written as follows:

$$\hat{Z}' = j\omega L_v(|1, Z_B\rangle\langle 1, Z_A'| + h.c.) + j\omega L_w(|2, Z_A'\rangle\langle 1, Z_B| + h.c.).$$
$$+j\omega L_v \sum_{m=2}^{N} (|m, Z_B\rangle\langle m, Z_A| + h.c.) + j\omega L_v \sum_{m=2}^{N} (|m, Z_B\rangle\langle m, Z_A| + h.c.) \quad (2)$$

Here, $Z_A'$ represents the impedance of loop 1 after the diode-induced symmetry breaking. We perform operations that involve chiral symmetry:

$$\hat{\Gamma}\hat{Z}\hat{\Gamma}^\dagger = -\hat{Z} \quad (3)$$
$$\hat{\Gamma}\hat{Z}'\hat{\Gamma}^\dagger \neq -\hat{Z}' \quad (4)$$

The above formula indicates that when the diodes conduct, it disrupts the periodic arrangement of hopping. The TMRM, initially in a topologically non-trivial state, transitions into a topologically trivial state. In the state of chiral symmetry breaking, by aligning the Larmor frequency with the topologically trivial state of the material, we effectively match the central magnetic field strength with that of the MR scanner's inherent coil (BC).

To verify that the TMRM's topological phase transition can effectively differentiate between the radiofrequency transmission field ($B_1^+$) and the signal field ($B_1^-$), MRI control experiments were also performed on a water phantom, with the setup essentially the same as the previously described experiments. The TMRM capable of topological phase transition and the FLC were placed on a fixed support (Supplementary Fig.9b). The enhancement of the RF signal field ($B_1^-$) by the material coil for the metamaterial surface is reflected in signal intensity, while the enhancement of the RF transmission field ($B_1^+$) is reflected in the flip angle. Images were captured using a $T_2$-weighted gradient-recalled echo sequence ($T_2$ GRE) sequence, with flip angles (FA) ranging from 10 to 90 degrees, using the BC as both the transmit and receive coil. The signal-flip angle curves of the experimental and the control group show the same trend, indicating that the TMRM's topological phase transition effectively eliminates the impact on the RF transmission field ($B_1^+$), as shown in Fig. 4b. TMRM significantly guides the RF signal field ($B_1^-$), while maintaining the RF transmission field ($B_1^+$) strength unchanged during the transmitting phase. This allows the TMRM to be suitable for both spin echo and gradient echo sequences without changing any parameters, ensuring convenience in clinical use.

**Conclusion**

In this work, we present a novel paradigm for MRI coil design and detail its underlying theory and methods. Utilizing topological circuits, we developed TMRM sheets that incorporate topological boundary states to significantly enhance MRI performance. Specifically, TMRM employs pseudo-two-dimensional dual topological boundary states to enable efficient signal transmission between the magnetic resonance signal and the MR scanner's receive coil. Simulations in CST Microwave Studio (Fig. 4c) demonstrate that when the TMRM is in a topologically non-trivial state, the central magnetic field strength is enhanced by a factor of 9.2 compared to the central magnetic field strength of the BC. Notably, the enhancement factor is significantly smaller than that of other metamaterial[21]. Moreover, TMRM exhibits topological phase transition capabilities that differentiate between the RF transmission field ($B_1^+$) and the signal field ($B_1^-$), thereby ensuring compatibility with both gradient echo (GE) and spin echo (SE) sequences while improving clinical convenience and accessibility.

TMRM serves as an effective complement to traditional array coils. By working in synergy with the built-in receive coils of the MR system, TMRM compensates for the inherent limitations of these coils—such as weak SNR and non-uniform field distribution. This collaborative integration not only enhances image SNR but also improves image uniformity and diagnostic accuracy. As TMRM augments the capabilities of conventional array coils, it offers a promising strategy to overcome current challenges in MR imaging technology.

In terms of imaging quality, TMRM improves the image signal-to-noise ratio (SNR), outperforming the standard four-channel flexible coil (4-ch FLC) and matching the performance of the specialized 12-channel wrist coil (12-ch WR). Additionally, TMRM offers substantial potential for configuration optimization; it can be fabricated in various shapes, such as elliptical or planar, to enhance conformability, increase the filling factor, and boost magnetic field strength (Supplementary Fig. 15).

Regarding parallel imaging, although TMRM operates in conjunction with local MR receive coils and reduces image artifacts through phase oversampling, its influence on the spatial sensitivity of these coils remains challenging to mitigate, resulting in interference (Supplementary Fig. 24). We anticipate that further application of PT theory will help minimize the impact of TMRM on local receive coils [51].

Despite the promising advantages, there are certain limitations that need to be addressed. In our preliminary demonstrations, the radial thickness of TMRM is approximately 5 cm. Due to the "body-boundary correspondence"[47], there are challenges in further optimizing the size and weight of TMRM. This size limitation could affect the adaptability of TMRM for certain anatomical regions or clinical applications where space is more constrained. Additionally, while TMRM improves parallel imaging by enabling oversampling, it still faces the drawback of interfering with the spatial sensitivity of local receive coils, which could limit its performance in certain scenarios.

Future work will further optimize this synergy to fully leverage the combined potential of TMRM and built-in receive coils in clinical applications. Additionally, a more comprehensive evaluation of TMRM's performance in clinical settings—particularly in comparison with multi-channel array receive coils and metamaterials that provide only localized enhancement—is crucial. This will allow us to better understand the advantages and limitations of metamaterials, as well as the unique characteristics and diagnostic value of the different mechanisms[27,32].

In summary, our findings highlight the promise of TMRM as a complementary technology that enhances MRI performance and paves the way for more reliable and versatile clinical imaging applications. Further optimization of TMRM's design and its integration with local receive coils will be key to overcoming current limitations and fully realizing its potential in clinical settings.

**Methods**
**1. Simulation**
The numerical solutions and operations of the SSH model in Fig. 2 were completed in MATLAB (MathWorks®). The solution of the topological circuit in Fig. 2 was done in LTspice® and plotted in MATLAB. Electromagnetic field simulations in Fig. 2-4 were performed in CST Microwave Studio and plotted in MATLAB. More details of simulation are provided in Supplementary S1.

**2. Fabrication of TMRM**
The TMRM units were designed and manufactured using commercial Printed Circuit Board (PCB) techniques. The substrate chosen was a copper base material, 70um thick, anti-oxidation Flexible Printed Circuit (FPC). Supplementary Fig.2 shows the specific size parameters of TMRM. For circuit layout and lumped elements, fixed capacitors produced by Dalian Dalicap Technology Co., Ltd. in China were selected (Supplementary Fig.3). Plastic molds for fixing the shape of the FPC were produced using 3D printing. The printed FPC was then fixed onto a 140*170mm, 60° bent plastic "V-shaped" mold and placed inside a cylindrical plastic structure to secure it. The inner diameter of the cylindrical plastic structure was 150mm, and the outer diameter was 180mm. Diodes used were PIN diodes from Infineon Technologies, model BAR63-04W. The fixed plastic support was created through 3D printing, with an outer diameter of 150mm and an inner diameter of 147mm.

**3. Experimental Setup**
The testing of the one-dimensional magnetic field distribution was performed with a custom-built planar magnetic field-testing system. This system included a vector network analyzer (Agilent N2530C), a magnetic field probe (LANGER mini magnetic field probe), a planar waveguide, and a stepper motor. MRI experiments were performed on the uMR570 1.5T MR system (United Imaging Healthcare). The water phantom solution was prepared by dissolving 2.6 g of NaCl and 0.3 g of $NiSO_4$ in every 1000 g of ultrapure water, then storing the prepared solution in a cylindrical container (used in Fig. 2-3). The cylinder had an inner diameter of 70 mm, an internal height of 120 mm, and a wall thickness of 3 mm. Cuboid water phantom was used in other experiments (Germany, Siemens, Supplementary Fig. 26). The fixed plastic support was positioned centrally within the MR scanner (Supplementary Fig. 6b).

**4. Participants**
There were 10 volunteers involved in the experiment. Five of them were male and five were female. The volunteers were randomly recruited from graduate students at the Tsinghua university. No sex or gender analysis was performed. Volunteers were placed in a prone position with their right arm placed in a TMRM (with the spine coil as receive coil), 4-ch FLC, or 12-ch WR (Supplementary Fig. 6a). During image acquisition, the TMRM was placed between the subject and spine coil, which was right above the examination bed.

All experiments involving humans and animals were approved by the Ethics Committee of Beijing Tsinghua Changgung Hospital, School of Medicine, Tsinghua University (Approval Number: 23306-0-01). Informed consent was obtained from all the participating volunteers.

**5. Image analysis**

The image signal-to-noise ratio is a single number obtained by dividing the image signal by the image noise. The calculation of SNR is performed according to the NEMA Standards Publication MS 1-2008 (R2014). Signal and noise are from the same MRI image. The signal is the mean signal within the region of interest (ROI), while the noise is the standard deviation of the signal in a background region located reasonably away from the edge of the sample. More details of simulation are provided in Supplementary S2-3.

**6. MRI scanning parameters**

MRI scanning parameters of Fig. 1 is in Supplementary Table2-3. MRI scanning parameters of Fig. 2 is in Supplementary Table4. MRI scanning parameters of Fig. 3 is in Supplementary Table5. MRI scanning parameters of Fig. 4 is in Supplementary Table6. All MRI sequences acquired in this study are 2D slice-selective. No normalization filter and image optimization techniques were active throughout the study.

**7. Standard Double Angle Method**

The standard dual-angle method (DAM) [52,53], is a common technique for measuring the radiofrequency (RF) field distribution in MRI, specifically the $B_1^+$ field intensity. This method involves performing two imaging sequences at the same location with different flip angles, allowing for the calculation of the actual $B_1^+$ field intensity.

DAM is based on the relationship between MR signal intensity and flip angle. By comparing MR images acquired with two different flip angles, $B_1^+$ field intensity can be derived. Typically, the chosen flip angles are θ and 2θ.

For a given flip angle θ, the MR signal intensity $S$ can be expressed as:
$$S(\theta) = M_0 \times \sin(\theta) \times E \tag{1}$$
where $M_0$ is the equilibrium magnetization, and $E$ is the decay factor related to the repetition time and the relaxation time.

For two different flip angles, θ and 2θ, the corresponding signal intensities $S_1$ and $S_2$ are given by:
$$S_1 = M_0 \times \sin(\theta) \times E \tag{2}$$
$$S_2 = M_0 \times \sin(2\theta) \times E = M_0 \times 2\sin(\theta)\cos(\theta) \times E \tag{3}$$

By comparing the two signal intensities, $S_1$ and $S_2$, the actual flip angle $\theta_{real}$ can be derived, allowing for the calculation of the $B_1^+$ field intensity.

Let the signal ratio $R$ be defined as:
$$R = \frac{S_2}{S_1} = 2\cos(\theta_{real}) \tag{4}$$
$$\theta_{real} = \arccos\left(\frac{R}{2}\right) \tag{5}$$

Since the $B_1^+$ field intensity is directly related to the flip angle $\theta_{real}$, the actual flip angle derived from the signal ratio can be used to calculate the $B_1^+$ field intensity distribution.

**6. Topological boundary states are also current resonance points**

According to the loop current method, the equation listing the loop current in Supplementary Fig.9 is as follows:

$$\begin{aligned} &\vdots \\ &2\int \frac{I_i dt}{C_i} + \left(L_{i-1}\frac{d}{dt} + R_{i-1}\right)(I_i - I_{i-1}) + \left(L_i\frac{d}{dt} + R_i\right)(I_i - I_{i+1}) = 0 \\ &2\int \frac{I_{i+1} dt}{C_{i+1}} + \left(L_i\frac{d}{dt} + R_i\right)(I_{i+1} - I_i) + \left(L_{i+1}\frac{d}{dt} + R_{i+1}\right)(I_{i+1} - I_{i+2}) = 0 \\ &\vdots \end{aligned} \tag{6}$$

Differentiate $t$ on both sides and organize it into a matrix form:
$$\bar{\bar{L}}\frac{d\bar{I}}{dt} + \bar{\bar{R}}\frac{d\bar{I}}{dt} + \frac{\bar{I}}{\bar{\bar{C}}} = 0, \qquad \bar{I} = (\cdots, I_i, I_{i+1}, \cdots) \tag{7}$$

$$\bar{\bar{L}} = \begin{pmatrix} \ddots & \cdots & 0 & 0 \\ \cdots & L_{i-1} + L_i & -L_{i-1} & 0 \\ 0 & -L_i & L_i + L_{i+1} & \cdots \\ 0 & 0 & \cdots & \ddots \end{pmatrix} \tag{8}$$

$\bar{\bar{R}}$, $\bar{\bar{C}}$ and so on.

$$\frac{Z(\omega)}{j\omega} = \left(\bar{L}j\omega + \bar{R} + \frac{1}{j\omega\bar{C}}\right) \tag{9}$$

Let $\dot{I} = \frac{dI}{dt}$, the preceding formula is reduced to

$$\frac{d}{dt}\begin{pmatrix}\dot{I}\\ I\end{pmatrix} = \begin{pmatrix}\bar{L}^{-1}\bar{R} & -\bar{L}^{-1}\bar{C}^{-1}\\ 1 & 0\end{pmatrix}\begin{pmatrix}\dot{I}\\ I\end{pmatrix} = \bar{A}\begin{pmatrix}\dot{I}\\ I\end{pmatrix} \tag{10}$$

The solution matrix is

$$e^{\bar{A}t} = e^{\lambda t}\sum_{i=0}^{n-1}\frac{t^i}{(i)!}(\bar{A} - \lambda E)^i \tag{11}$$

$\lambda$ is the n-double eigenvalue of $\bar{A}$.

$$\det(\bar{A} - \bar{\lambda}E) = 0 \tag{12}$$

$$\det[(\bar{L}^{-1}\bar{R} + \lambda)\lambda + \bar{L}^{-1}\bar{C}^{-1}] = 0 \tag{13}$$

Let $\lambda = j\omega$

$$\det[Z(\omega)] = 0 \tag{14}$$

It is proved that when the circuit in Supplementary Fig.9 is in the topological boundary state, the current of the circuit is in the resonant state

**Reporting Summary**. Further information on research design is available in the Nature Research Reporting Summary linked to this article.

## Data availability
The main data supporting the results in this study are available within the paper and its Supplementary Information. Anonymized MRI data are available in Zenodo at https://doi.org/10.5281/zenodo.13756448.

## Code availability
The DICOM image viewing software used is uVIEWER (United Imaging, China). LTspice schematics and MATLAB codes for the DAM, H-field and SSH model are available in GitHub at https://github.com/2771743166/TMRM.git.

**Acknowledgements**
This work is supported by the National Key Research & Development Program of China (No. 2023YFB3811400), National Natural Science Foundation of China (No. 52273296), Beijing Municipal Science & Technology Commission (No. Z221100006722016). The authors are deeply grateful to Li He, Deqing Jiang, Yichen Wang, and Yue Jin for their assistance with the experiments.

**Author contributions**
Ji Zhou and Qian Zhao designed the study. Siyong Zheng and Maopeng Wu designed and performed all the experimental measurements and drafted the paper. Zhonghai Chi, Xinxin Li, Mingze Weng, Fubei Liu, Yingyi Qi, and Zhuozhao Zheng analyzed the results. All authors contributed to scientific discussion and the final version of the manuscript.

**Competing interests**
The authors declare no competing financial interests.

## Supplementary Information (SI).

### S1. Simulation setup in CST

Numerical simulations are performed using the frequency-domain solver of CST Microwave Studio 2020 package. In open boundary situation, the simulation is excited with the circularly polarized plane wave. To acquire the magnetic field distribution, an H-Field monitor is set at the interested frequency. When the TMRM is placed in body coil (birdcage coil, BC), the simulation is excited with two discrete ports orthogonally placed in space and 90° difference in the phase (Supplementary Fig. 1a). The BC was tuned to 63.8 MHz. The structural parameters of TMRM are detailed in Supplementary Fig. 2. The lumped elements are arranged on the TMRM (Supplementary Fig. 1b). The simulation model replicated the dimensions of the fabricated sample. The cylindrical substrate, which has no influence on electromagnetic wave, is eliminated in the simulation. $B_1^+$, $B_1^-$, and SAR are normalized to 1.0 W accepted power.

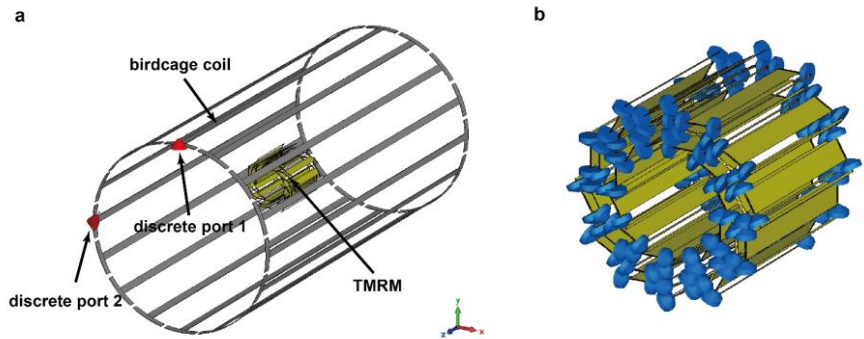

**Supplementary Fig. 1. (a)** Simulation model in CST software, the TMRM is placed in body coil. **(b)** Simulation model in CST software, the lumped elements (the blue arrows) are placed on the TMRM.

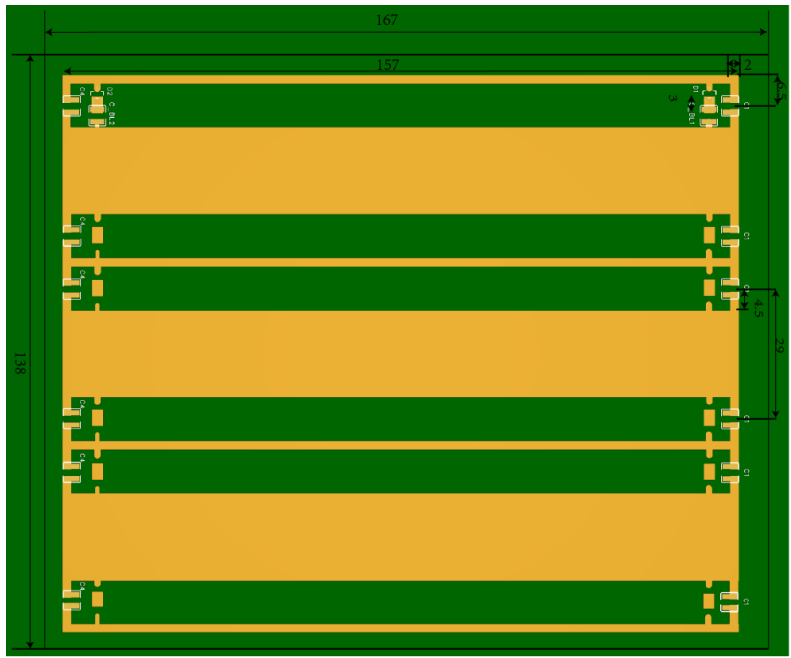

**Supplementary Fig. 2.** Structural parameters of simulation and fabrication of topological metamaterial structural units (unit: mm).

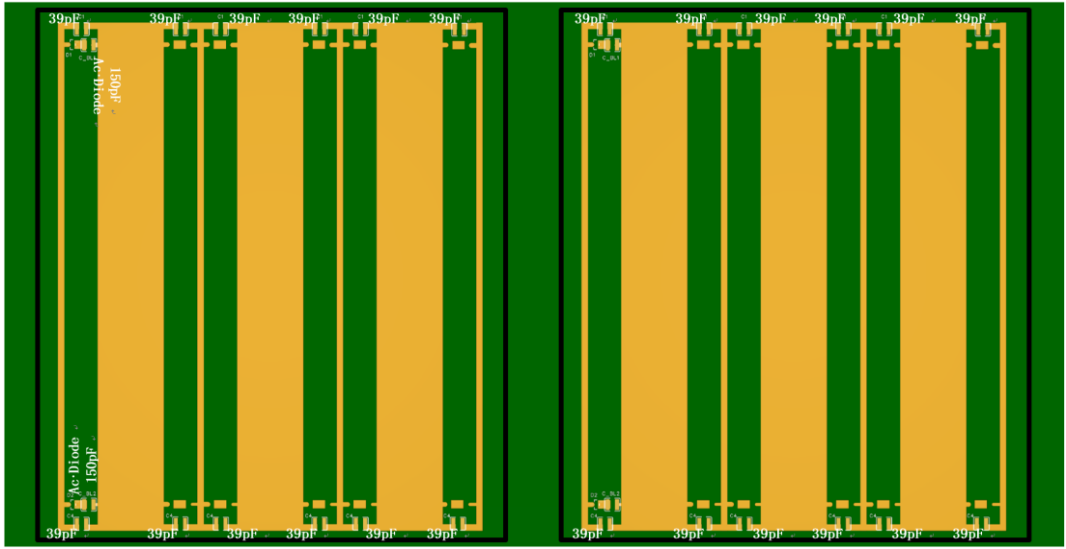

**Supplementary Fig. 3.** Lumped element parameters for simulation and fabrication of topological metamaterial structural units.

**S2. Image analysis**

The calculations of SNR of the MR images were performed based on the National Electrical Manufactures Association Standards Publication MS 1-2008 (R2014).

Image SNR analysis of wrist joint MRI scans of volunteers and water phantom: SNR = signal/noise. Signal and noise are from the same MRI image. The signal is the mean signal within the region of interest, whereas the noise was measured as the SD of the signal in a region that is reasonably distant from the edge of the sample in the background. The regions of interest of different tissue and noise of wrist joint were marked in Supplementary Fig. 4. Following the Supplementary Fig. 4, SNR data from 10 volunteers (including signals within ROI and noise at image edges) are presented in supplementary table 7-13. The regions of interest and noise of water phantom were marked in Supplementary Fig. 4.

For uniformity, maximum (Smax) and minimum (Smin) signal values were measured in a centered and regular geometric area enclosing at least 75% of the image of the signal. Image uniformity = 100 × (1−(Smax−Smin)/(Smax + Smin)). The ROI preference is consistent with the image SNR analysis of the water phantom in Supplementary Fig. 5.

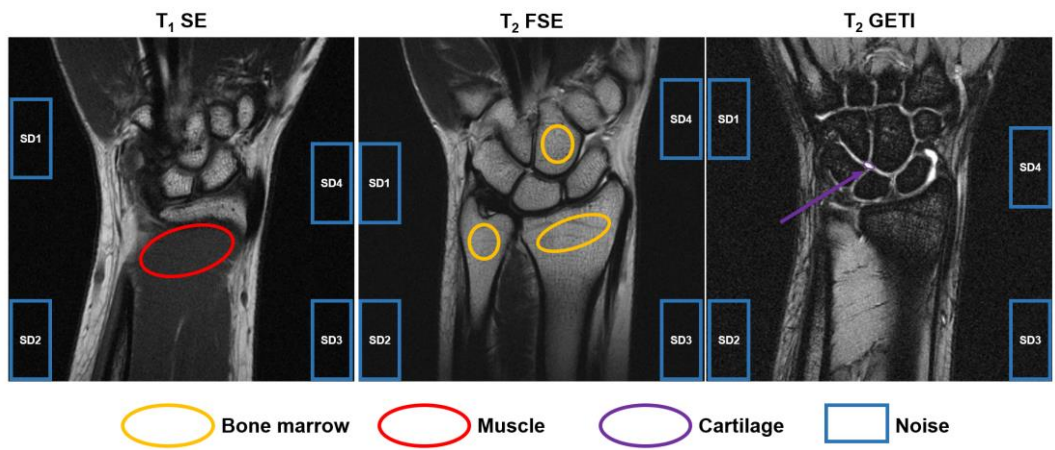

**Supplementary Fig. 4.** Regions of interest in SNR analysis of the wrist joint. Different types of tissue are marked by different colours of circles. Red is muscle, purple is cartilage, orange is bone marrow, and blue is where noise is measured.

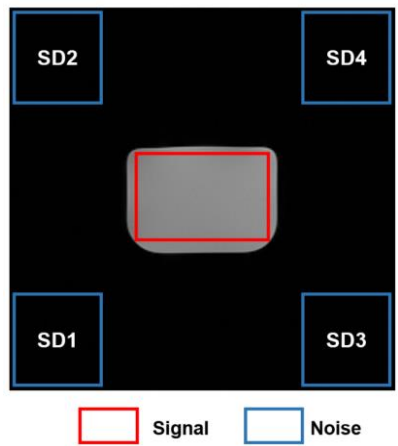

**Supplementary Fig. 5.** Regions of interest in SNR analysis of the water phantom. Red is signal, and blue is where noise is measured.

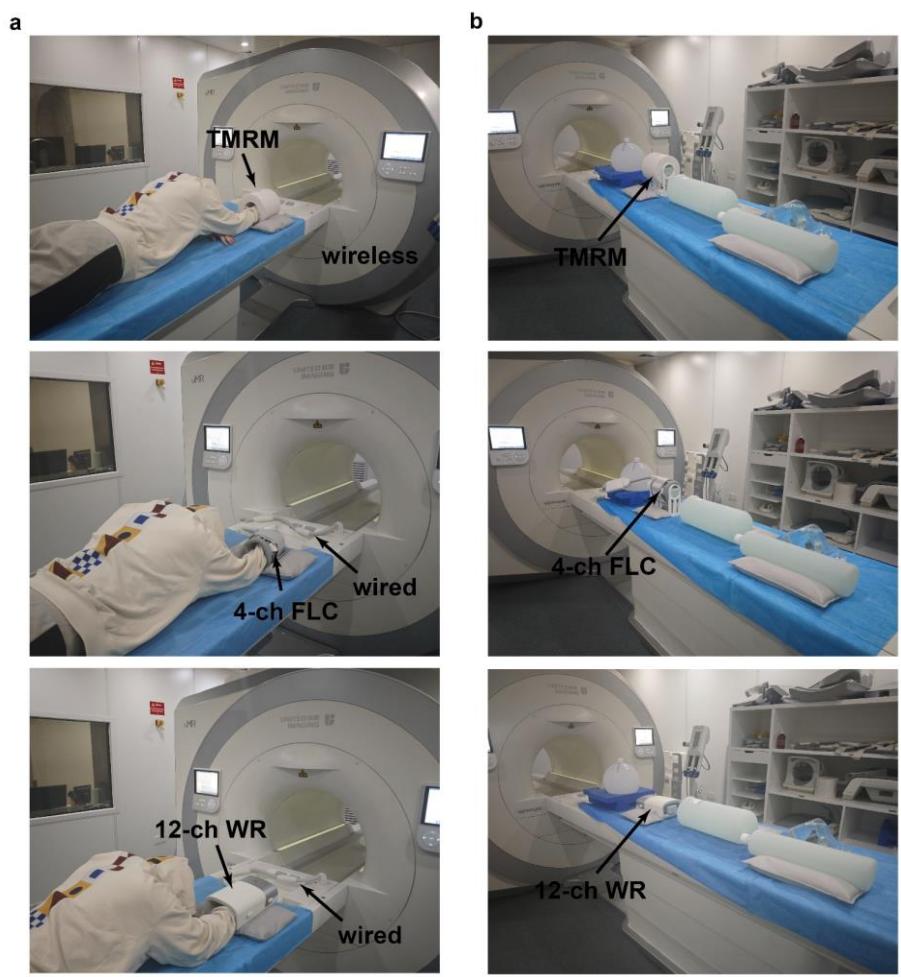

**Supplementary Fig. 6. (a)** The setup for volunteer testing. **(b)** The setup images of the water phantom imaging experiment. The MR scanner is the uMR570 1.5 T MR scanner (United Imaging, China). The receive coils used in volunteer testing is a Spine array receive coil (SP) and in the water phantom imaging experiment is a birdcage body coil (BC).

**S3. Statistical analysis**

Data are presented as the mean ± SEM and analysed using OriginPro (Learning Edition). The application, Paired Comparison Plot, is downloaded to OriginPro (https://www.originlab.com/fileExchange/details.aspx?fid=390). This app can be used to create various plot with significant differences. Multiple comparisons were tested using the two-way analysis of variance (ANOVA). Tukey's multiple comparisons test was used for the adjustment of p-value threshold.

Image quality was assessed independently by 2 experienced radiologists (at least 5 years of experience) using a 5-point Likert scale. All MR images were reviewed by two radiologists with at least 5 y of experience (Y.Y. and Y.W.) in clinical musculoskeletal MR imaging. The images were reviewed independently, and the final decisions reached by consensus were reported. The reviewers were blinded to the coil information during the MRI image interpretations. For the coronal MRI images of human wrists, image quality was scored from 1-5. 5: the bone, tendon, muscle, and joint compartment are clearly shown without image artifacts; 4: the bone, tendon, muscle, and joint compartment are clearly shown with slight image artifacts, which do not affect diagnosis; 3: part of the structure is blurred, which does not affect the important structures; 2: most of the structures are vague and may affect the diagnosis; 1: the structure of the wrist is unclear and the diagnosis cannot be made.

**S4. SAR analysis**

To estimate the SAR of the TMRM in a MR scanner, we constructed a high-pass BC in simulation (Supplementary Fig. 1a), serving as both the transmitter and receiver. The BC was tuned to 63.8 MHz with the TMRM positioned at its isocenter. In simulation, we replaced the PIN diodes with good conductor. We adjust the capacitance in series with the PIN diodes to achieve the desired frequency drift, thus determining the optimal conditions for reception and transmission phases. $B_1^+$, $B_1^-$, and SAR are normalized to 1.0 W accepted power. The permittivity of the water phantom is 78, the conductivity is 0.52 S/m, and the density is 1000 kg/m$^3$.

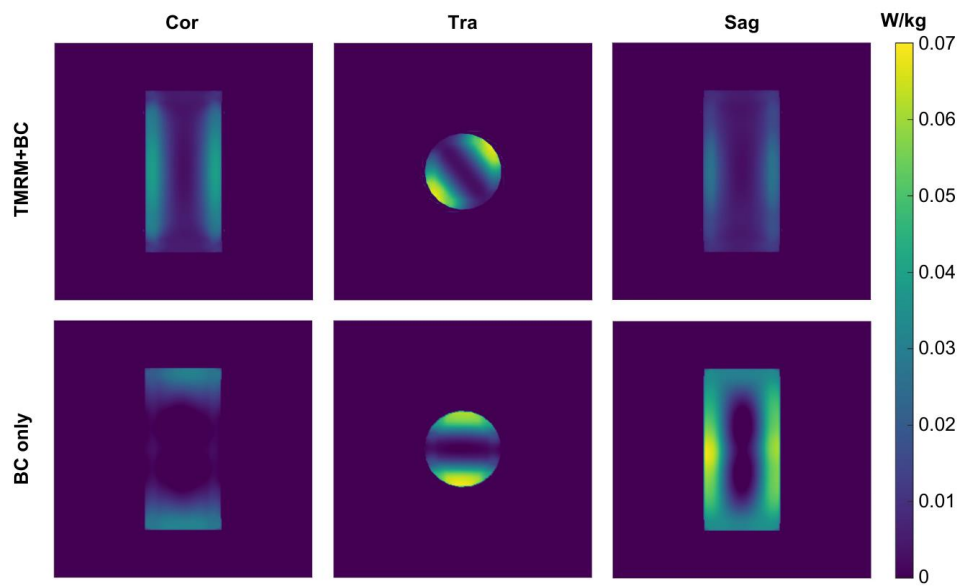

**Supplementary Fig. 7.** SAR of TMRM and BC. Max SAR of TMRM+BC is 0.073 W/kg. Max SAR of BC is 0.071 W/kg.

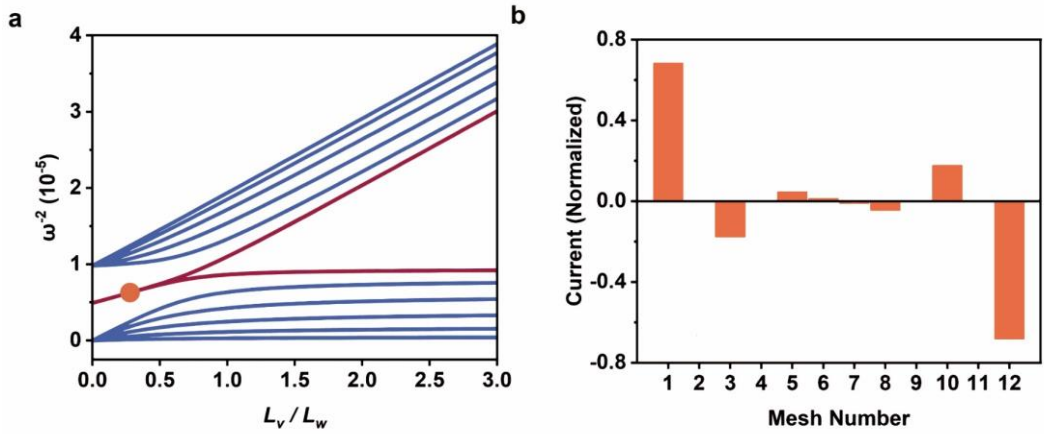

**Supplementary Fig. 8.** (a) The graph depicts the energy spectrum's evolution as a function of the ratio between strong (intracell) and weak (intercell) coupling. It highlights a transition from a topological phase with edge states ($L_v < L_w$) to a trivial phase without in-gap states ($L_v > L_w$), showcasing the impact of coupling strength on the system's topology. (b) The current distributions of two end states.

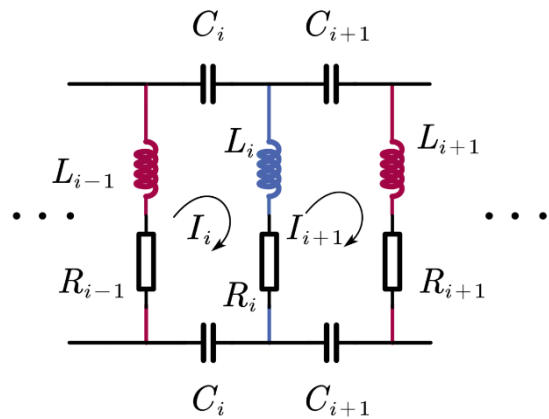

**Supplementary Fig. 9.** Equivalent circuit of topological magnetic resonance metamaterial unit

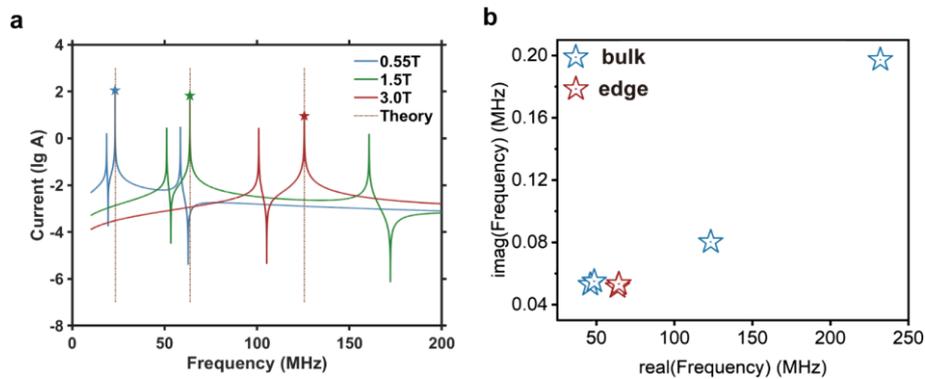

**Supplementary Fig. 10.** (a) Frequency response diagram of the TMRM current from SPICE simulation. (b) Diagram of the imaginary and real parts of the frequencies of various modes of the TMRM.

## S5. The influence of stacking

There is coupling between TMRM sheets. Bending the TMRM sheet can make the magnetic field distribution evolve from ITBS to DTBS (Supplementary Fig. 11). In the main (Fig.3), the angle of DTBS is 30 degrees, and the angle of ITBS and ETBS is 0 degrees.

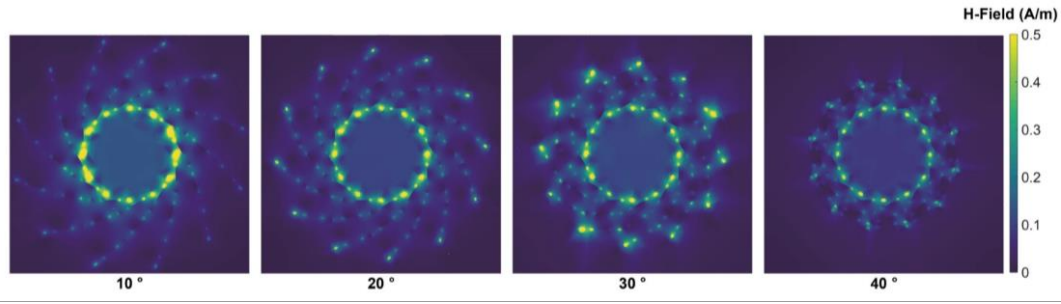

**Supplementary Fig. 11.** Influence of bending angle of TMRM sheet on magnetic field distribution.

## S6. The influence of angle relative to the magnet axis

In open boundary conditions, we simulate the H-Field of the TMRM to the angle relative to the magnet axis (Supplementary Fig. 12a). Due to the difference in TMRM to the angle relative to the magnet axis, the central magnetic field strength of TMRM will vary (Supplementary Fig. 12b). The first case to consider in the simulation is $k_p=0$. As the TMRM gradually rotates, the central magnetic field strength of the TMRM gradually decreases, and the uniformity of the magnetic field is basically unchanged. This is because TMRM is designed to be excited by circularly polarized waves. However, there is coupling between TMRM sheets, so a uniform magnetic field distribution can be maintained under on-line polarization excitation.

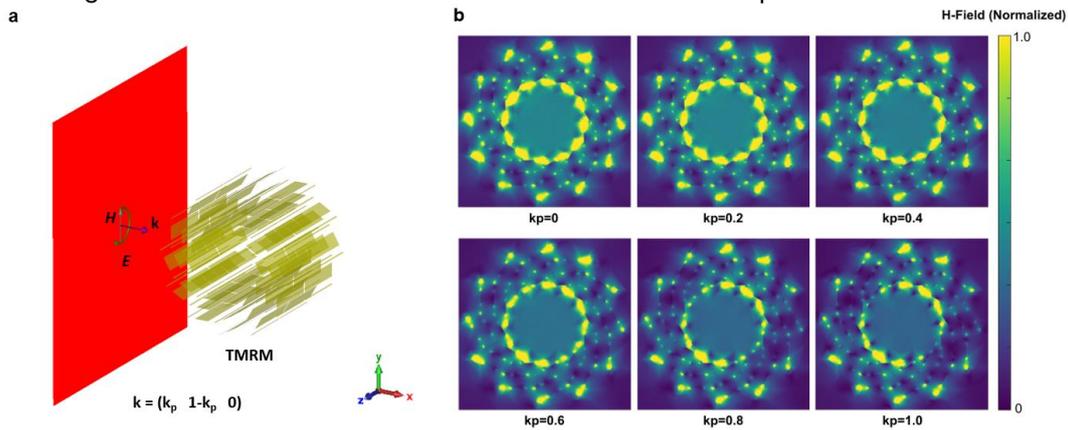

**Supplementary Fig. 12.** H-Field of the TMRM to the angle relative to the magnet axis. **(a)** The setup of simulation of the TMRM to the angle relative to the magnet axis. **(b)** H-Field of TMRM to the angle relative to the magnet axis.

## S7. The concept of chiral symmetry

A chiral phenomenon in physics is one that is not identical to its mirror image. The word chirality came to be used to refer directly to the transformation property, such that spinors transforming one way are said to be "right-handed" or of "positive chirality", and those transforming the other way are said to be "left-handed" or of "negative chirality". For instance, the wave function $\psi$ in the Dirac representation can be written as $\psi = \begin{pmatrix} \psi_L \\ \psi_R \end{pmatrix}$, the transformation laws for left-handed spinor $\psi_L$ and right-handed spinor $\psi_R$, under infinitesimal rotations $\theta$ and boots $\beta$, are

$$\psi_L \to \left(1 - \frac{i}{2}\theta \cdot \sigma - \frac{i}{2}\beta \cdot \sigma\right)$$
$$\psi_R \to \left(1 - \frac{i}{2}\theta \cdot \sigma + \frac{i}{2}\beta \cdot \sigma\right)$$

Among various symmetries, chiral symmetry plays important roles in various condensed matter systems. It becomes important especially when considering simple tight-binding models. We say that a systems with tight-binding Hamiltonian $H$ has chiral symmetry, if $\hat{\Gamma} H \hat{\Gamma}^\dagger \neq -H$ with $\hat{\Gamma}$ being the chiral symmetry operator. In such systems, chiral symmetry is often attributed to bipartite lattice structures, thus is often called the sublattice symmetry as well, i.e., the symmetry operation that changes the sign of wave functions on all sites of one of the two sublattices of the bipartite lattice. To illustrate that, we can define orthogonal sublattice projectors $P_A$ and $P_B$, as $P_A = \frac{I+\hat{\Gamma}}{2}$ and $P_B = \frac{I+\hat{\Gamma}}{2}$, where $I$ represents the identity operator on the Hilbert space of the system.

## S8. Control circuit and $B_1^+$ of TMRM

During the transmitting period, AC diode is conducted. During the receiving period, AC diode is disconnected.

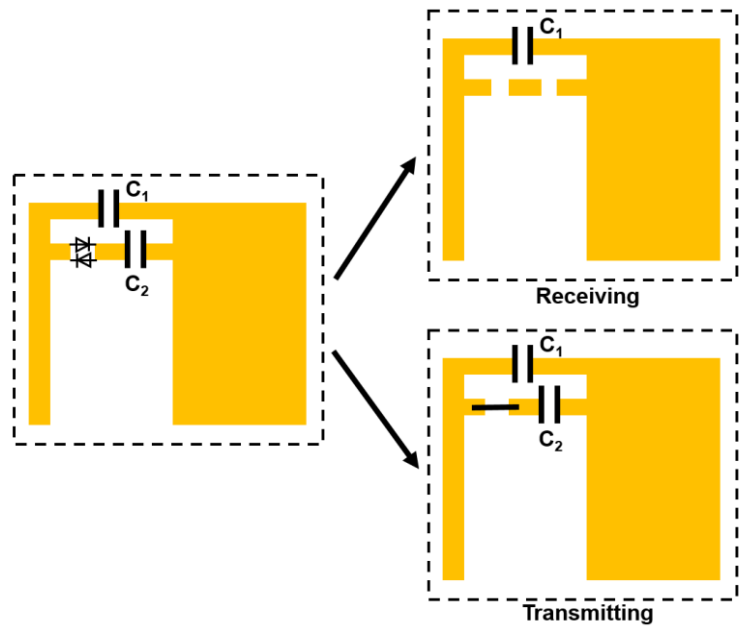

**Supplementary Fig. 13.** Control circuit state of the TMRM in the transmitting and receiving periods.

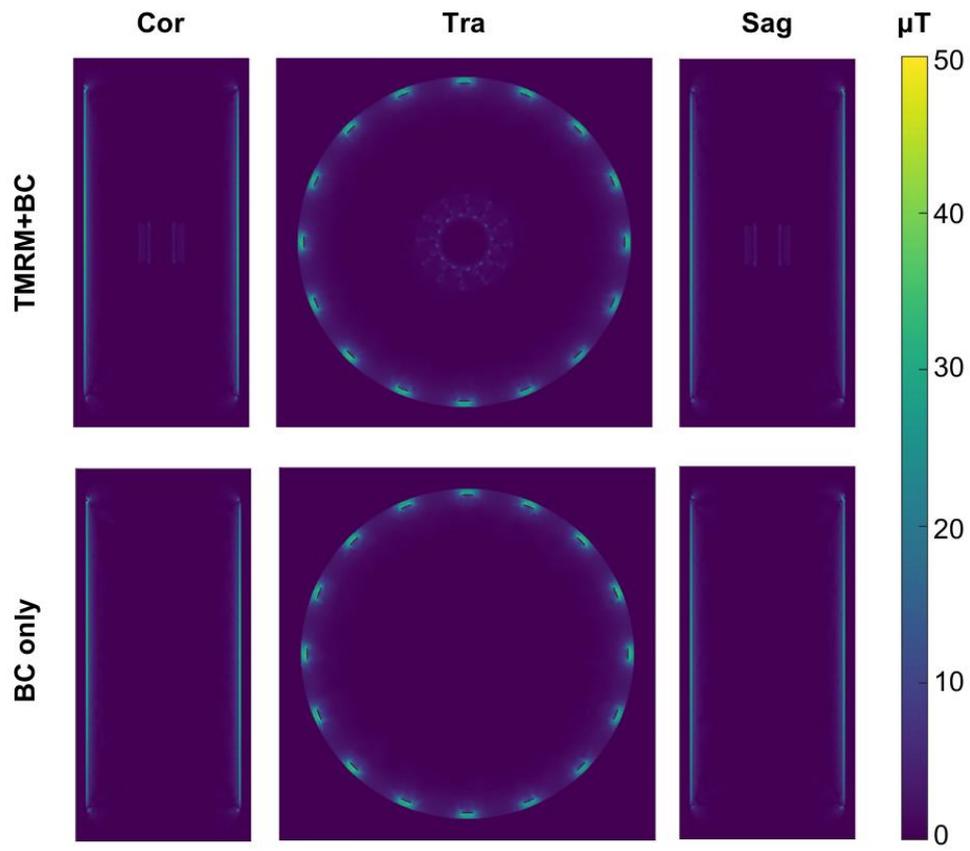

**Supplementary Fig. 14.** $B_1^+$ of TMRM and BC.

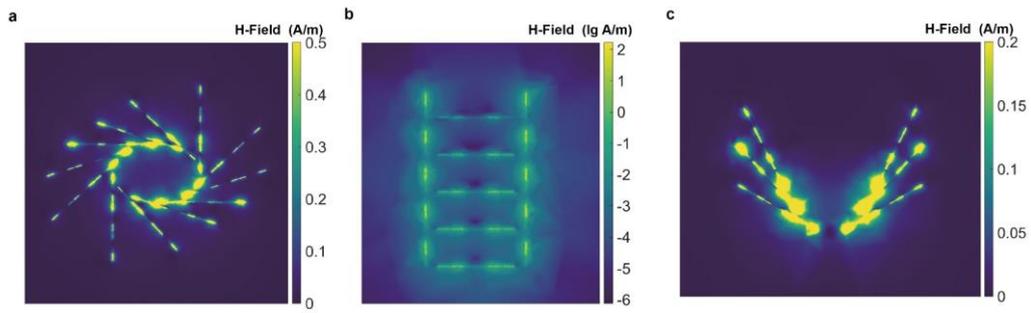

**Supplementary Fig. 15.** H-field of 3 configurations of TMRM. **(a)** Further optimization of the conformal elliptical configuration. **(b)** The "staple" configuration is suitable for flat anatomical regions. **(c)** The curved configuration is designed for small-area imaging.

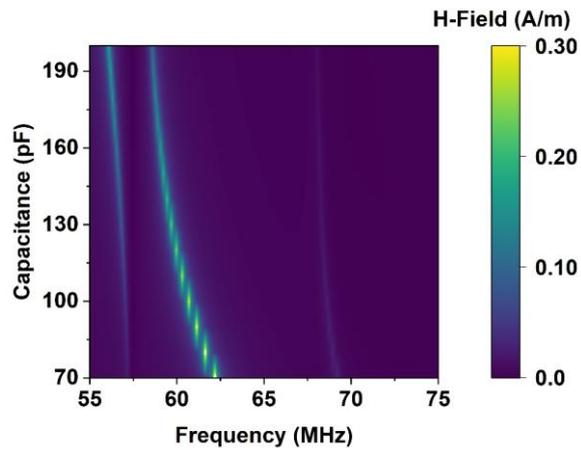

**Supplementary Fig. 16.** H-field vs. frequency of TMRM at different capacitance.

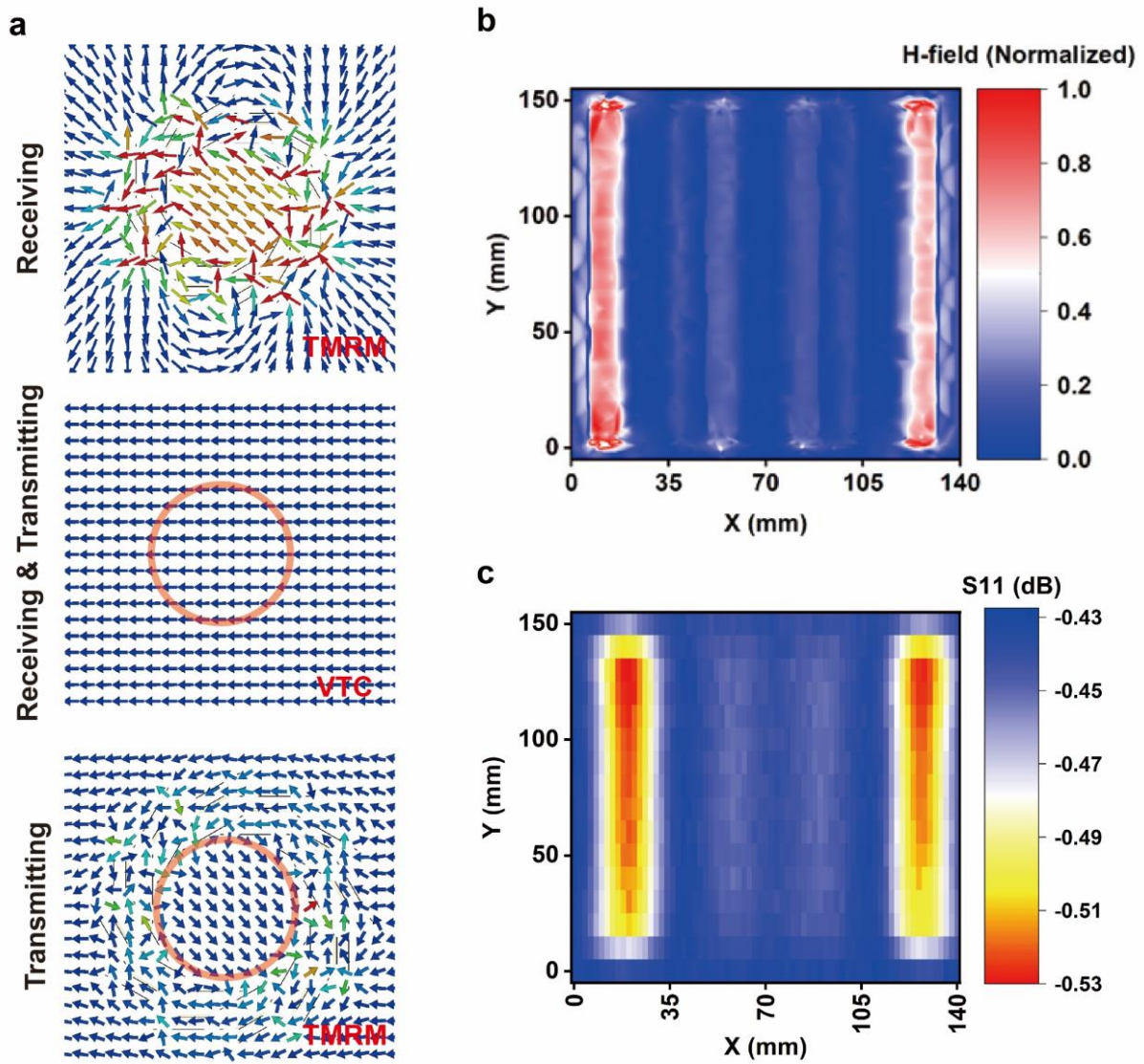

**Supplementary Fig. 17.** **(a)** Phase of the $B_1^+$ fields in the ROI. **(b)** A simulation diagram of the one-dimensional magnetic field distribution of the TMRM unit. **(c)** A test image of the one-dimensional magnetic field distribution of an actual fabricated TMRM unit. The TMRM unit exhibits the same topological boundary states as the theoretical design.

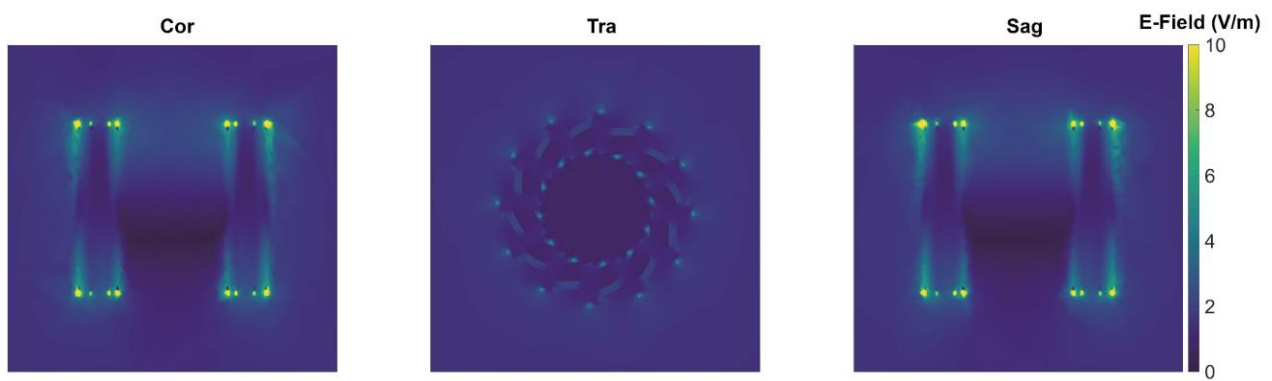

**Supplementary Fig. 18.** E-Field of TMRM in receiving phase. In open boundary situation, the simulation is excited with the circularly polarized plane wave.

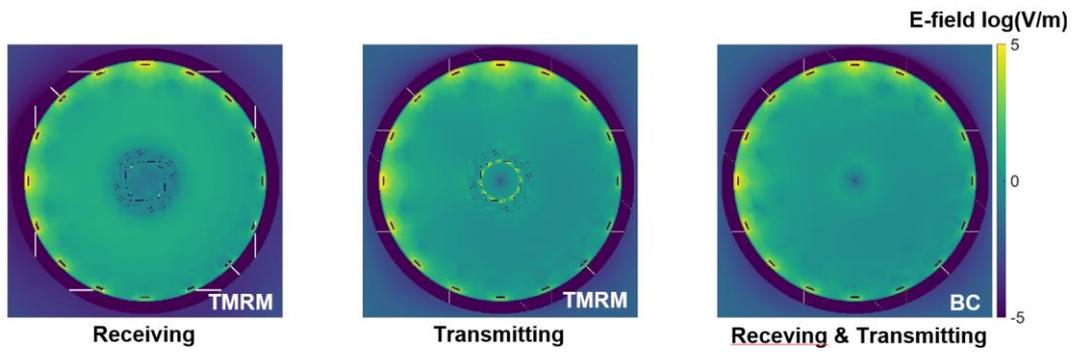

**Supplementary Fig. 19.** E-Field simulation illustrating the topological phase transition caused by maintaining and breaking chiral symmetry protection.

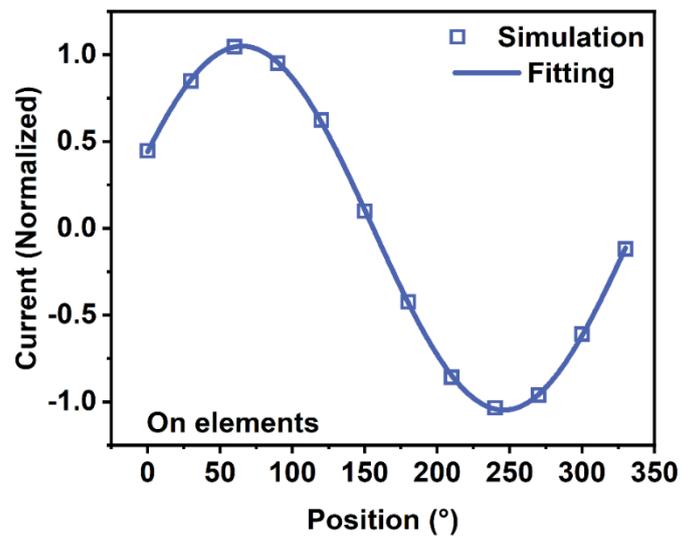

**Supplementary Fig. 20.** Simulation Image of Current Distribution on the Circumference of the TMRM.

**S8. Information about 4-ch FLC, 12-ch WR and TMRM**

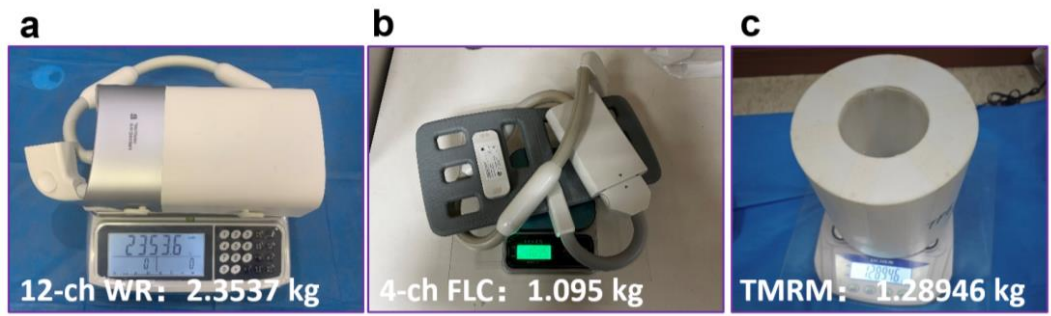

**Supplementary Fig. 21. Weight of 12-ch WR and TMRM. (a)** Weight of 12-ch WR is 2.35kg. **(b)** Weight of 4-ch FLC is 1.10kg. **(c)** Weight of TMRM is 1.29kg.

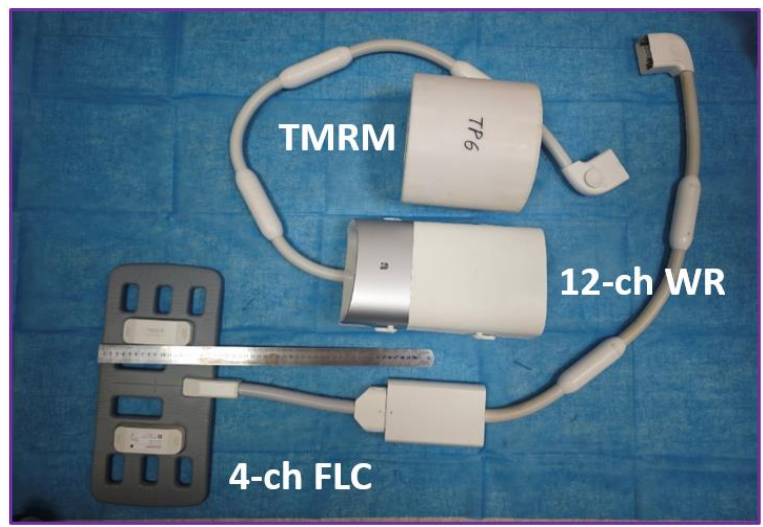

**Supplementary Fig. 22.** Photo of 4-ch FLC, 12-ch WR and TMRM.

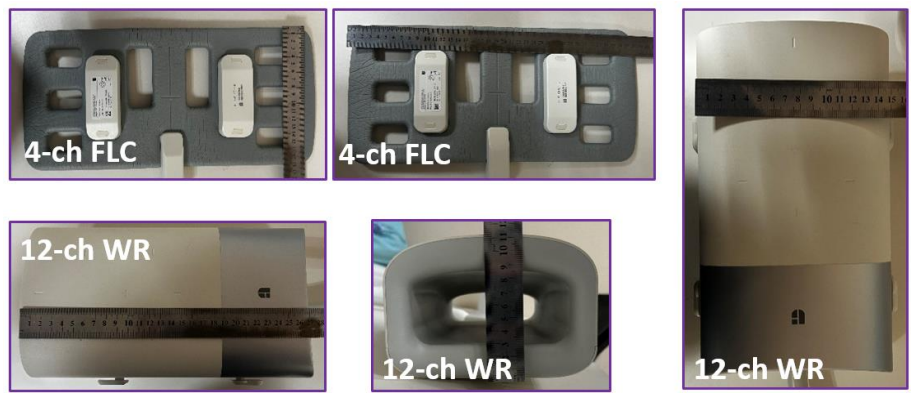

**Supplementary Fig. 23.** Size of 4-ch FLC, 12-ch WR and TMRM. The dimensions of the 4-ch FLC (United Imaging Healthcare, China) are 380 mm × 174 mm. It is a wrap-around coil made from soft and flexible material, featuring four integrated low-noise preamplifiers. The dimensions of the 12-channel dedicated wrist coil (12-ch WR, uMR570, United Imaging Healthcare, China) are 160 mm × 270 × 110 mm. It is designed for imaging of wrist or hand.

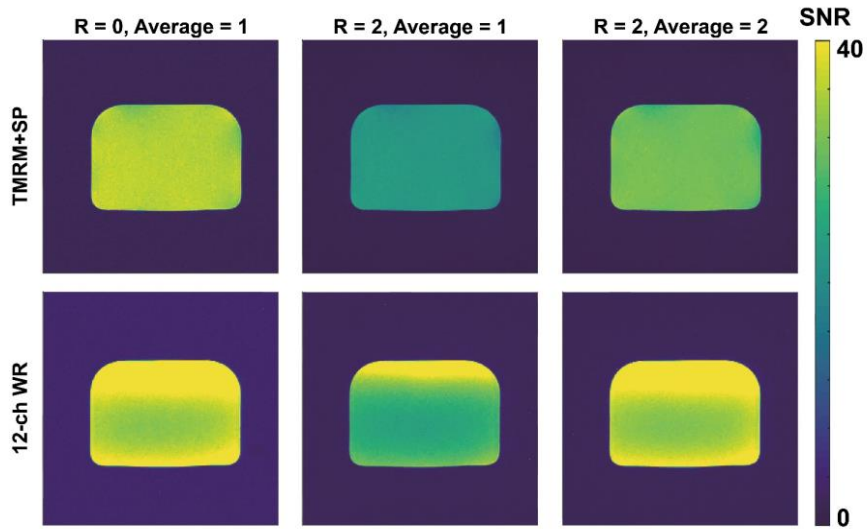

**Supplementary Fig. 24.** SNR maps of 12-ch WR and TMRM on water phantom. The MRI sequence parameters for $T_2$ FSE are as follows: the repetition time (TR) is 5000 ms, the echo time (TE) is 88.62 ms, and the flip angles are 90° and 150°. The field of view (FOV) is 100×100 mm, with a slice thickness of 2.5 mm and in-plane resolution of 0.39×0.39 mm. Phase oversampling is set to 100%, and the acceleration direction is from right to left.

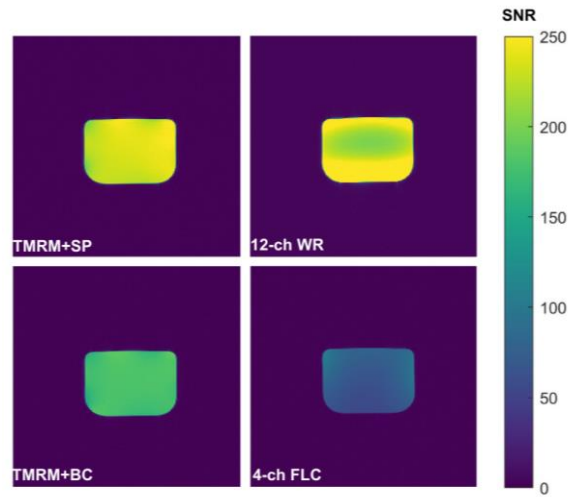

**Supplementary Fig. 25.** SNR maps of 4-ch FLC, 12-ch WR and TMRM on water phantom. The MRI sequence parameters for SE are presented in Supplementary Table3.

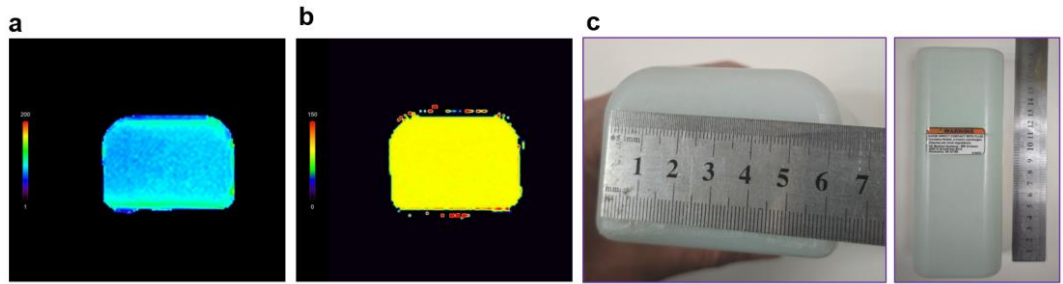

**Supplementary Fig. 26. Cuboid water phantom. (a)** $T_1$ Map of Water phantom. **(b)** $T_2$ Map of Water phantom. **(c)** Size of water phantom.

**Supplementary Table1.** The inductance values of the metal structures in the TMRM units.

| Self-induction/Mutual induction | Value (nH@64MHz) |
|:---:|:---:|
| $L_v$ | 65.44 |
| $L_w$ | 117.1 |
| $M_1$ | 30.00 |
| $M_2$ | 13.97 |
| $M_2'$ | 15.18 |

**Supplementary Table 2.** MRI scanning parameters of human wrist joint (Fig. 1)

| Sequence | TR | TE | ROI/mm$^2$ | SD/mm | Flip Angle/° |
|---|---|---|---|---|---|
| T$_1$ SE | 420.0 | 15.1 | 100*100 | 0.39*0.39*2.50 | 90/180 |
| T$_2$ FSE | 5000.0 | 88.8 | 100*100 | 0.39*0.39*2.50 | 90/150 |
| T$_2$ GETI | 828.9 | 25.5 | 100*100 | 0.39*0.39*2.50 | 30 |

**Supplementary Table 3.** Parameters of Water Phantom MRI scanning (Fig. 1)

| Sequence | TR | TE | ROI/mm$^2$ | SD/mm | Scanning time/min | Flip Angle/° |
|---|---|---|---|---|---|---|
| SE | 1200.0 | 30 | 160*160 | 0.63*0.63*5.00 | 5:14 | 90/180 |

**Supplementary Table 4.** Parameters of water phantom MRI scanning (Fig. 2)

| Sequence | TR | TE | ROI/mm$^2$ | SD/mm | Flip Angle/° |
|---|---|---|---|---|---|
| SPGR | 500.0 | 7.0 | 500*500 | 0.98*0.98*1.80 | 4/8 |

**Supplementary Table 5.** Parameters of water phantom MRI scanning (Fig. 3)

| Sequence | TR | TE | ROI/mm$^2$ | SD/mm | Flip Angle/° |
|---|---|---|---|---|---|
| T$_2$ GETI | 762.8 | 31.0 | 100*100 | 0.39*0.39*2.00 | 40 |

**Supplementary Table 6.** Parameters of water Phantom MRI scanning (Fig. 4)

| Sequence | TR | TE | ROI/mm$^2$ | SD/mm |
|---|---|---|---|---|
| T$_2$ GRE | 500.0 | 25.0 | 100*100 | 0.78*0.78*5.00 |

**Supplementary Table7.** $T_1$ SE: SNR of bone marrow

| Volunteer | TMRM | 4-ch FLC | 12-ch WR |
| --- | --- | --- | --- |
| volunteer 1 | 23.0 | 9.0 | 24.3 |
| volunteer 2 | 24.6 | 9.1 | 26.2 |
| volunteer 3 | 24.4 | 7.9 | 29.5 |
| volunteer 4 | 21.6 | 9.4 | 23.6 |
| volunteer 5 | 24.1 | 8.9 | 26.9 |
| volunteer 6 | 20.4 | 8.8 | 27.6 |
| volunteer 7 | 25.4 | 8.4 | 26.8 |
| volunteer 8 | 27.2 | 9.3 | 29.2 |
| volunteer 9 | 24.7 | 8.2 | 24.2 |
| volunteer 10 | 19.8 | 8.1 | 27.8 |

**Supplementary Table8.** $T_1$ SE: SNR of muscle

| Volunteer | TMRM | 4-ch FLC | 12-ch WR |
| --- | --- | --- | --- |
| volunteer 1 | 9.9 | 3.6 | 11.1 |
| volunteer 2 | 11.3 | 2.3 | 9.9 |
| volunteer 3 | 11.3 | 3.5 | 20.8 |
| volunteer 4 | 8.1 | 3.6 | 11.0 |
| volunteer 5 | 10.4 | 3.0 | 10.2 |
| volunteer 6 | 10.4 | 3.5 | 11.0 |
| volunteer 7 | 9.2 | 3.6 | 11.2 |
| volunteer 8 | 11.6 | 3.8 | 13.2 |
| volunteer 9 | 9.5 | 3.5 | 10.3 |
| volunteer 10 | 8.8 | 3.5 | 10.4 |

**Supplementary Table9.** $T_2$ FSE: SNR of bone marrow

| Volunteer | TMRM | 4-ch FLC | 12-ch WR |
| --- | --- | --- | --- |
| volunteer 1 | 25.6 | 11.3 | 28.4 |
| volunteer 2 | 30.2 | 8.9 | 32.3 |
| volunteer 3 | 29.1 | 9.7 | 34.2 |
| volunteer 4 | 25.4 | 9.8 | 25.3 |
| volunteer 5 | 28.3 | 9.8 | 32.7 |
| volunteer 6 | 25.6 | 10.4 | 31.7 |
| volunteer 7 | 25.4 | 9.3 | 30.0 |
| volunteer 8 | 27.0 | 9.4 | 28.2 |
| volunteer 9 | 30.6 | 10.9 | 35.3 |
| volunteer 10 | 24.4 | 10.2 | 28.8 |

**Supplementary Table10.** $T_2$ FSE: SNR of muscle

| Volunteer | TMRM | 4-ch FLC | 12-ch WR |
| --- | --- | --- | --- |
| volunteer 1 | 6.9 | 2.9 | 8.5 |
| volunteer 2 | 9.6 | 3.0 | 10.4 |
| volunteer 3 | 7.2 | 2.8 | 8.8 |
| volunteer 4 | 7.9 | 3.0 | 9.0 |
| volunteer 5 | 9.1 | 3.2 | 10.4 |
| volunteer 6 | 7.6 | 2.9 | 9.6 |
| volunteer 7 | 8.8 | 2.9 | 10.8 |
| volunteer 8 | 8.1 | 3.0 | 8.6 |
| volunteer 9 | 7.3 | 3.4 | 10.4 |
| volunteer 10 | 6.0 | 2.6 | 7.7 |

**Supplementary Table 11.** T$_2$ GETI: SNR of bone marrow

| Volunteer | TMRM | 4-ch FLC | 12-ch WR |
|---|---|---|---|
| volunteer 1 | 7.4 | 3.9 | 9.4 |
| volunteer 2 | 9.1 | 3.7 | 10.2 |
| volunteer 3 | 8.6 | 3.9 | 10.7 |
| volunteer 4 | 7.4 | 3.9 | 8.9 |
| volunteer 5 | 8.7 | 4.0 | 9.9 |
| volunteer 6 | 7.7 | 3.9 | 10.1 |
| volunteer 7 | 9.0 | 3.8 | 10.3 |
| volunteer 8 | 8.4 | 3.8 | 9.4 |
| volunteer 9 | 9.9 | 4.2 | 10.6 |
| volunteer 10 | 6.9 | 3.9 | 10.2 |

**Supplementary Table 12.** T$_2$ GETI: SNR of muscle

| Volunteer | TMRM | 4-ch FLC | 12-ch WR |
|---|---|---|---|
| volunteer 1 | 18.7 | 5.8 | 18.9 |
| volunteer 2 | 21.6 | 5.9 | 17.3 |
| volunteer 3 | 20.4 | 5.9 | 31.3 |
| volunteer 4 | 20.5 | 6.2 | 14.2 |
| volunteer 5 | 21.3 | 6.0 | 19.6 |
| volunteer 6 | 18.1 | 5.4 | 18.2 |
| volunteer 7 | 19.4 | 6.1 | 17.4 |
| volunteer 8 | 23.4 | 8.8 | 17.4 |
| volunteer 9 | 18.3 | 5.2 | 16.9 |
| volunteer 10 | 18.9 | 7.0 | 16.2 |

**Supplementary Table13.** $T_2$ GETI: SNR of cartilage

| Volunteer | 12-ch WR | 4-ch FLC | TMRM |
|---|---|---|---|
| volunteer 1 | 15.5 | 4.6 | 16.2 |
| volunteer 2 | 21.4 | 4.7 | 20.8 |
| volunteer 3 | 23.8 | 6.8 | 25.2 |
| volunteer 4 | 27.5 | 7.8 | 16.6 |
| volunteer 5 | 20.1 | 4.7 | 19.4 |
| volunteer 6 | 6.3 | 5.7 | 12.2 |
| volunteer 7 | 15.5 | 4.9 | 13.8 |
| volunteer 8 | 24.2 | 6.0 | 16.0 |
| volunteer 9 | 21.0 | 6.9 | 18.9 |
| volunteer 10 | 15.0 | 5.8 | 16.2 |